\documentclass[preprint,3p,times,authoryear]{elsarticle}

\usepackage{graphicx}
\usepackage{amsmath}
\usepackage{amssymb}
\usepackage{amsfonts}
\usepackage{mathrsfs}
\usepackage{adjustbox}
\usepackage{mdframed}
\usepackage[figuresright]{rotating}

\usepackage[utf8]{inputenc}
\usepackage[T1]{fontenc}
\usepackage[final]{changes}

\usepackage{lipsum}

\usepackage{dutchcal}

\graphicspath{ {figures/} }

\usepackage{nomencl}
\usepackage{etoolbox}
\usepackage{wrapfig}
\usepackage{framed} 
\usepackage[skins,breakable, most]{tcolorbox}

\renewcommand*\nompreamble{\begin{multicols}{2}}
\usepackage{textgreek}

\usepackage{epsfig}
\usepackage{epstopdf} 
\usepackage{todonotes}
\usepackage{setspace}
\usepackage{siunitx}
\usepackage{tabularx}
\usepackage{longtable} 

\usepackage{verbatim}

\usepackage[caption=false]{subfig}
\usepackage[hyphens]{url}
\usepackage{graphicx}
\usepackage{booktabs}

\usepackage{lmodern,textcomp}
\usepackage{rotating} 
\usepackage{multicol} 
\usepackage{multirow} 
\usepackage{mathtools, cuted}
\usepackage{lipsum}
\usepackage{tabularx} 
\newcolumntype{L}[1]{>{\raggedright\arraybackslash}p{#1}} 
\newcolumntype{C}[1]{>{\centering\arraybackslash}p{#1}} 
\newcolumntype{R}[1]{>{\raggedleft\arraybackslash}p{#1}} 
\usepackage{amssymb}
\usepackage{pifont}
\RequirePackage[ngerman=ngerman-x-latest]{hyphsubst}
\usepackage{microtype}
\usepackage[american]{babel} 
\usepackage{hyperref}
\usepackage{float}
\usepackage[normalem]{ulem}
\usepackage{colortbl}
\useunder{\uline}{\ul}{}
\usepackage{lscape}
\usepackage{marvosym}

\usepackage{amsmath, amssymb}

\usepackage{lineno}

\biboptions{square,comma,sort&compress}

\setcitestyle{numbers}
\journal{}
\makeatletter
\def\ps@pprintTitle{%
 \let\@oddhead\@empty
 \let\@evenhead\@empty
 \def\@oddfoot{}%
 \let\@evenfoot\@oddfoot}
\makeatother
\usepackage{longtable}  
\usepackage{tabulary}

\usepackage[acronym]{glossaries}
\usepackage{glossary-inline}
\renewcommand*\nompreamble{\begin{multicols}{2}}
\makeglossaries
\newacronym{eu}{EU}{European Union}
\newacronym{ghg}{GHG}{Greenhouse gases} 
\newacronym{ptg}{PtG}{Power-to-Gas}
\newacronym{ptx}{PtX}{Power-to-X}
\newacronym{ptl}{PtL}{Power-to-Liquids}
\newacronym{pth2}{PtH2}{Power-to-H2}
\newacronym{re}{RE}{Renewable energy}
\newacronym{res}{RES}{Renewable energy sources}
\newacronym{ccs}{CCS}{Carbon capture and storage}
\newacronym{BMWi}{BMWi}{Federal Ministry for Economic Affairs and CLimate Protection (Bundesministerium für Wirtschaft und Klimaschutz)}
\newacronym{eeg}{EEG}{Renewable Energy Sources Act (Erneuerbare-Energien-Gesetz)}
\newacronym{iea}{IEA}{International Energy Agency}
\newacronym{fchju}{FCH JU}{Fuel Cells and Hydrogen Joint Undertaking}
\newacronym{H2}{H2}{Hydrogen}
\newacronym{ipcc}{IPCC}{Intergovernmental Panel on Climate Change}
\newacronym{uba}{UBA}{Federal Environment Agency (Umweltbundesamt)}
\newacronym{bau}{BAU}{Business-as-ususal}
\newacronym{gw}{GW}{Gigawatts}
\newacronym{pv}{PV}{Photovoltaic}
\newacronym{lulucf}{LULUCF}{Land Use, Land-Use Change and Forestry}
\newacronym{CO2e}{CO2e}{CO2 equivalents}
\newacronym{ptprate}{PTP rate}{Pure time preference rate}
\newacronym{nrw}{NRW}{North Rhine-Westphalia}
\newacronym{bw}{BW}{Baden-Wuerttemberg}

\newcommand\blfootnote[1]{%
 \begingroup
 \renewcommand\thefootnote{}\footnote{#1}%
 \addtocounter{footnote}{-1}%
 \endgroup
}

\begin{document}

\begin{frontmatter}


\title{Future role and economic benefits of hydrogen and synthetic energy carriers in Germany: a systematic review of long-term energy scenarios}


\author[DTU,IIRM]{Fabian Scheller\corref{cor1}}
\ead{fjosc@dtu.dk}
\cortext[cor1]{Corresponding author}
\author[IIRM]{Stefan Wald}
\author[IIRM]{Hendrik Kondziella}
\author[DTU]{Philipp Andreas Gunkel}
\author[IIRM]{Thomas Bruckner}
\author[DTU]{Dogan Keles}

\address[DTU]{Energy Economics and System Analysis, DTU Management, Technical University of Denmark (DTU), Denmark}
\address[IIRM]{Institute for Infrastructure and Resources Management (IIRM), Leipzig University, Germany}

\begin{abstract}
Determining the development of Germany's energy system by taking the energy transition objectives into account is the subject of a series of studies. Since their assumptions and results play a significant role in the political energy debate for understanding the role of hydrogen and synthetic energy carriers, a better discussion is needed. This article provides a comparative assessment of published transition pathways for Germany to assess the role and advantages of hydrogen and synthetic energy carriers. Twelve energy studies were selected and 37 scenarios for the years 2030 and 2050 were evaluated. Despite the variations, the two carriers will play an important future role. While their deployment is expected to have only started by 2030 with a mean demand of 91 TWh/a (\textasciitilde4\% of the final energy demand) in Germany, they will be an essential part by 2050 with a mean demand of 480 TWh/a (\textasciitilde24\% of the final energy demand). A moderately positive correlation (0.53) between the decarbonisation targets and the share of hydrogen-based carriers in final energy demand underlines the relevance for reaching the climate targets. Additionally, value creation effects of about 5 bn EUR/a in 2030 can be expected for hydrogen-based carriers. By 2050, these effects will increase to almost 16 bn EUR/a. Hydrogen is expected to be mainly produced domestically while synthetic fuels are projected to be mostly imported. Despite of all the advantages, the construction of the facilities is associated with high costs which should be not neglected in the discussion.

\end{abstract}

\begin{keyword}
Hydrogen \sep Synthetic fuels \sep Energy system analysis \sep Energy transition \sep Economic benefits \sep Meta study \end{keyword}

\end{frontmatter}
\section*{Highlights}
\begin{itemize}
\item Economic benefits of hydrogen and synthetic carriers are determined for Germany based on 37 scenarios.
\item Hydrogen and synthetic demand vary widely and average 91 (480) TWh/a for 2030 (2050).
\item Share of hydrogen-based carriers in final energy demand correlates positively with emissions reductions.
\item Value creation effects of 5 bn EUR/a for 2030 and 16 bn EUR/a for 2050 are calculated.
\item Germany's strategic goal of the years 2030 seem quite ambitious and additional efforst might be needed.

\end{itemize}

\blfootnote{\printglossary[type=\acronymtype,style=inline,nonumberlist]}


\section{Introduction}
\label{S:1}

\subsection{Hydrogen and synthetic fuels as integral part of the future energy system}
At the global scale, hydrogen will play a significant role in reaching net-zero emissions within the next three decades. Given ambitious scenario targets, its share in total final energy consumption is expected to increase from 0.2\% (2020) to 10\% (2050) \cite{IEA.2021}. In the European countries a twice as high share is even expected \cite{FCHJU.2019}. One reason could be that in 2019 the European Commission plans to become the first continent to achieve climate neutrality by 2050. The so-called “European Green Deal” proposes multiple measures to promote investments into green technologies, to support industry and economy in the energy transition, to decarbonise the energy sector, to increase energy efficiency and strengthen international cooperation on environmental standards \cite{EU2020}. The recent accentuation of European climate policy is also coinciding with the national emission reduction targets of, e.g., Germany \cite{BMWi.2021}, and the development of hydrogen technologies with low-carbon potential is in line with ambitious climate targets. Governmental announcements from leading industrial states all over the globe and policy programs supporting hydrogen-based technologies underline this development in recent years \cite{IEA.2019}. 

While politicians are in relative consensus about the targets to be set, it is still an open question what an adequate energy system of the future will look like and what energy carriers and technologies will be implemented to achieve the goals \cite{Szarka.2017}. One indication could be delivered by the existing scenario and modelling studies examining this question and project different transition pathways towards a future sustainable energy system. Future energy scenarios can be helpful to evaluate developments in the energy sector and thus provide a basis for discussion. That way, they can support the political decision-making process \cite{Dieckhoff.2014}. At the same time, single studies can hardly provide robust statements about future developments. The scenarios themselves do not attempt to predict the future but rather to examine the technical feasibility of defined targets and to provide a benchmark for the future development of the energy system \cite{FraunhoferISE2020b}. In this context, this paper focuses on the integration of various future scenarios of the German energy transition (c.f., \cite{Prognos.2020b,dena.2018b}) to get more reliable insights. The reason for the chosen country is twofold. On the one hand, Germany pursue the goal of being essentially greenhouse gas neutral in the year 2045 \cite{BMWi.2019}. Many of the conclusions and pathways encountered in this paper are also be transferable to other demand-intensive countries. In addition Germany follows a a special strategy follows with the phase-out of nuclear power until 2022 and the exit from coal-fired power until 2038 \cite{BMWi.2019,GWS.2018}. Thus, the results could be a preprint also for long-term intentions of other industrial states.

While the generation capacity of \acrlong{re} (\acrshort{re}) depends on the exact configuration of the energy system, the expansion is emphasised as one crucial element of the energy transition \cite{Prognos.2020b,dena.2018a}. The increasing share of intermittent \acrlong{res} (\acrshort{res}) comes along with several challenges. These include grid stability, as wind energy, in particular, is generated to a large extent in the north of Germany and has to be transported to the south, which creates imbalances; sufficient storage capacity to serve demand when no energy can be provided by wind or solar energy; and high flexibility to ensure the integration of \acrshort{res} \cite{Markard2018}. Hydrogen might play a pivotal role in the integration of \acrshort{re}, and the implementation of sector coupling \cite{Fraunhofer2019}. This is also underlined by the national hydrogen strategy \cite{BMWi.2020}. 

As a versatile energy carrier, hydrogen offers applications and can be used in various sectors. It will be particularly important as a raw material in the industry and as an energy carrier in the transport sector, especially in applications that cannot be electrified, e.g. shipping and aviation. In addition, it enables the long-term storage of electricity in terms of chemical energy \cite{BMWi.2020} and thus securing the energy supply in times when, e.g., \acrshort{res} are not available \cite{Bunger.2016}. Climate-friendly produced hydrogen (green hydrogen\footnote{Most studies and the federal government regard only ‘green hydrogen’ produced by water electrolysis (using \acrshort{re}) as sustainable. This view is adopted in this paper. In contrast, some studies (c.f, \cite{FCHJU.2019}) also consider a scenario with a similar CO\textsubscript{2} abatement potential where hydrogen is produced by methane steam reforming followed by carbon capture and storage (\acrshort{ccs}) (‘blue hydrogen’).}) thus contributes to decarbonising the energy system. Hydrogen can be further processed and converted into synthetic energy carriers in gaseous or liquid forms, which is referred to as \acrlong{ptx}  (\acrshort{ptx}) or \acrlong{ptl} (\acrshort{ptl}) or \acrlong{ptg} (\acrshort{ptg}), respectively. Although the terms are used differently by the different studies, in the following the term \acrshort{ptx} refers to all synthetic energy products in gaseous form, including hydrogen. Thereby, \acrshort{ptx} comprises in this paper carriers which are otherwise described with \acrshort{ptg}) or \acrshort{ptl}. However, due to the multiple conversion steps, synthetic energy carriers have the disadvantage of a relatively low overall efficiency for the entire process. They also suffer from high production costs, which impedes further market penetration \cite{Schnuelle.2019}. Fazeli et al. \cite{Fazeli.2021} emphasise that only the introduction of a growing CO\textsubscript{2} price will lead to hydrogen being able to compete with fossil fuels in terms of price. Similarly, Schnuelle et al. \cite{Schnuelle.2019} point out that higher carbon or EU ETS prices would make \acrshort{ptg} products more competitive.


The large-scale introduction of hydrogen solutions depends on various factors \cite{Hanley.2018}. In addition to ambitious climate targets and a high share of \acrshort{re} in the energy system as main conditions, these include an available hydrogen infrastructure and the question of whether economical solutions can be found in sectors where emissions avoidance is difficult. Hanley et al. \cite{Hanley.2018} see it as the responsibility of policy-makers to provide suitable support measures. According to the Federal Ministry of Economics and Climate Protection \cite{BMWi.2019}, as of today, hydrogen has no impact on the energy supply of Germany and is mainly used for substance recovery. The share of green hydrogen is relatively small, with 3 TWh out of 55 TWh of, primarily grey, produced hydrogen \cite{BMWi.2019}. Currently, green-based hydrogen cannot substitute grey hydrogen because higher production costs concerning the electrolysis are not economically viable \cite{gas20192030}. Referring to \cite{bukold2020blauer}, production costs of green hydrogen are between 15 and 18 ct/kWh, whereas blue hydrogen has production costs of 6.3 ct/kWh and grey hydrogen 4.5 ct/kWh on average \cite{bukold2020blauer}. The most dominant parameters of production costs for green hydrogen are the electrolyser’s electricity costs and capital expenditures.

The industrial relevance gained importance in recent years. The funding from the German government alone amounts to 9 billion euros \cite{BMWi.2020}. This large amount raises the question of whether the promotion is appropriate. Studies on the effects of the energy transition in general as well as investigations on economic value added and employment in the energy sector have been conducted more frequently (c.f., \cite{Hirschl.2010,Prognos.2015}). With regard to the economic effects from the production, storage and transport of hydrogen, studies investigate its potential in Germany \cite{LBST.2019} or certain regions of Germany (e.g., in \acrfull{nrw} \cite{LBST.2019} or in \acrfull{bw} \cite{McKenna.2018}), evaluate the advantages and disadvantages of imports compared to domestic production \cite{WuppertalInstitut.2020}, while others focus on applications like fuel cell buses \cite{Coleman.2020}. Regarding the calculations, LBST \cite{LBST.2019} determined the value creation from hydrogen production between 12 bn EUR and 50 bn EUR per year until 2050, with the largest share (45\% to 75\%) coming from the installation of renewables. Fraunhofer ISI and Fraunhofer ISE \cite{Fraunhofer2019} stated 32 bn for value creation the year 2050. Frontier Economics and IW Köln \cite{frontiereconomics.2018} even calculate value creation effects for the German industry of 36 bn EUR annually until 2050. In addition to the positive economic effects through employment and value creation, hydrogen can also make a “significant contribution” to the reduction of GHG emissions \cite{WuppertalInstitut.2020}. All in all, Acar and Dincer \cite{Acar.2019} list eight different contributions of hydrogen, each of which can be assigned to either the decarbonisation or integration aspect which also strengthen the resilience of the energy system \cite{Dincer.2016,Acar.2019}. This papers investigates investigates whether some of these advantages are also applicable to a highly industrialised country like Germany.  


\subsection{Research scope and objectives}

While many energy system studies and scenarios exist that examine the future of the energy system, the investigated objectives and the underlying assumptions of each study can be very different. This makes it difficult to assess the results generally. Meta-studies are a proper method to get a comprehensive overview of results from different studies.
Samadi et al. \cite{Samadi.2018}, e.g., compare the \cite{BCG.2018} study with other scenario studies to “gain insights for the assessments and discussions of current energy policy goals” \cite{Samadi.2018}. Existing meta-analyses of the German future energy system focus on a specific aspect, such as biomass \cite{Szarka.2017}, sector coupling or \acrshort{ptg} \cite{Forschungsradar.2018, Ecofys.2018, AEE.2016}. Adelphi, Wuppertal Institute, Dena, and IEEJ \cite{adelphi.2019} conduct a comprehensive meta-analysis of eighteen scenarios from six studies, in which they comprehensively examine both the future energy system and the role of hydrogen in it. At the same time, the economic benefits offered by the energy transition in general and the introduction of hydrogen and other synthetic carriers were neglected. One of few studies that use data from energy system studies on hydrogen demand as the basis for their calculations of value-added and job effects is the study by Wuppertal Institute, and DIW Econ \cite{WuppertalInstitut.2020}. However, synthetic fuels were not taken into account in this study. Additionally, Wiese et al. \cite{wiese2022strategies} analyses the the imports of hydrogen, methane and synthetic fuels energy carriers regarding the future German energy systems. Global reviews of the role of hydrogen and synthetic fuels were conducted by Chapman et al. \cite{chapman2019review} and Hanley et a. \cite{Hanley.2018}.

Thus, we want to answer the following question with the paper at hand: what are the economic benefits of the implementation of hydrogen into the energy system for Germany by 2030 and 2050? In order to answer the research question systematically the objectives of this paper are twofold. First, a comparative analysis of a large number of energy scenarios focusing on the future role of hydrogen and synthetic energy carriers in Germany is provided. While the focal energy carriers are hydrogen and synthetic energy carriers, we included different national energy studies with different objectives. Second, the future demand for hydrogen and hydrogen-based products projected in the studies is evaluated from a macroeconomic perspective. The value creation effects resulting from the introduction of hydrogen and other synthetic energy carriers are examined. The results give an indication of the benefits of a future energy system, which includes hydrogen and \acrshort{ptg} as a component, from a macroeconomic perspective. A reference to hydrogen subsidy payments can thus be established. An overview of the general research design is outlined in Figure \ref{fig:StudiesOverview}.

\begin{figure}[ht!]
    \centering
    \includegraphics[width=0.7\textwidth]{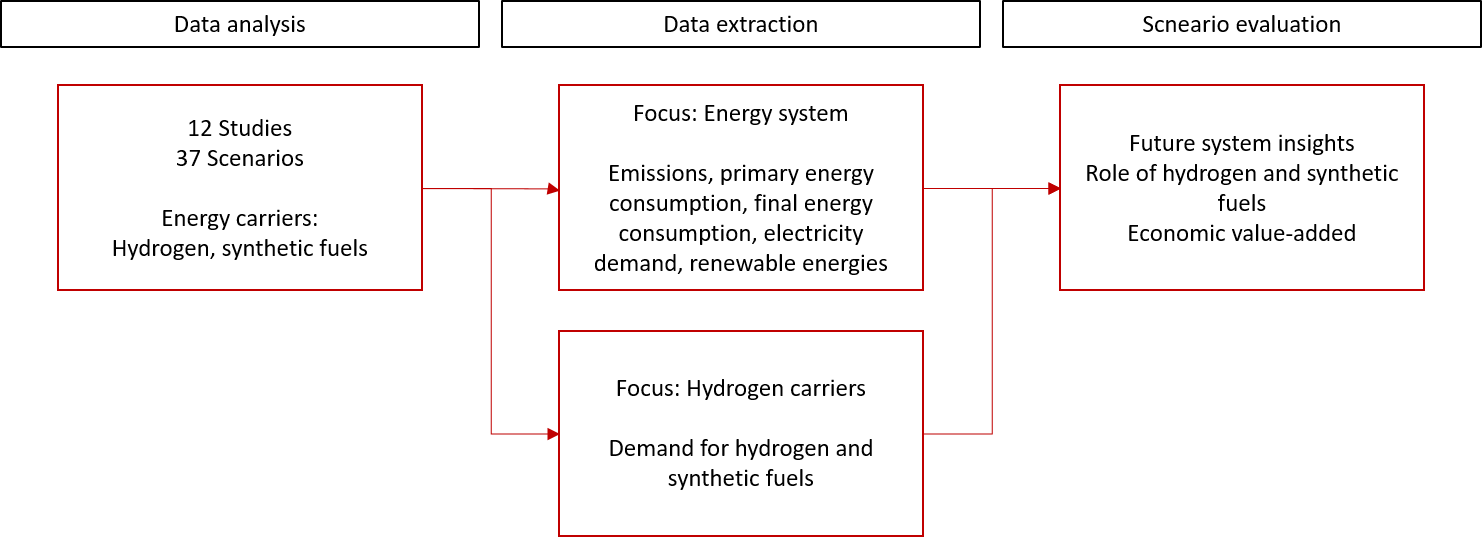}
    \caption{Overview of main research design with respect to the analysis procedure and research objective. We analysed 12 energy system studies and evaluated the role of hydrogen.}
    \label{fig:StudiesOverview}
\end{figure}

\section{Methodology}
\label{S:2}

\subsection{Scenario selection and data extraction}\label{sec:datacollection}

A literature review was conducted to identify energy system studies with long-term scenarios to set up the meta-database for our assessments. For enhanced comparability of the studies and the data, the following selection criteria were applied: first, the studies had to deal with the German energy system; second, only studies published in the year 2017 or later were considered due to the updated expansion plans for \acrshort{res}; third, the selected studies had to cover the target year 2050; and fourth, studies that focused on the entire energy system or/and the carrier hydrogen were within the scope of this paper. 

As a result, we selected a total of 37 scenarios from 12 studies as outlined in Table \ref{tab:studies}. Nearly all studies analysed the entire energy system beyond hydrogen. An exception represents the study of Fraunhofer ISI, and ISE \cite{Fraunhofer2019} which only examines the hydrogen aspect. Nevertheless, the studies vary in their comprehensiveness and level of detail. Some studies, like the one by Dena \& ewi \cite{dena.2018a} or the one by Prognos and \"Oko-Institut and Wuppertal-Institut \cite{Prognos.2020b} analyse the entire energy system very precisely while others focus on specific aspects. For instance, the study by Frontier Economics, AEW, 4Management, and EMCEL \cite{frontiereconomics.2017} concentrate on the role of the German gas infrastructure, and the study of Fraunhofer ISE \cite{FraunhoferISE2020a} puts the focus on the implications of social behaviour for the energy transition. 

Furthermore, the studies usually perform a scenario analysis that is one of the following types: (1) it is assumed that policies, technologies and preferences in society will continue to evolve similarly and at a similar rate as in the past (\acrfull{bau}); (2) a target scenario concerning a clear goal (e.g., the achievement of climate neutrality by the year 2050) is set and plausible pathways to reach it are investigated. Concerning the selected studies, most scenarios describe transformation pathways that are in line with the target corridor set by the German government in its previous climate protection plan of minus 80\% to minus 95\% GHG emissions by 2050 compared to 1990 levels \cite{BMU.2016}. A reduction of 55\% was targeted as a milestone until 2030 and was adopted by various studies, while others do not model this interim target. The dena Baseline, BDI Reference and ewi RF scenarios are \acrshort{bau} scenarios. They have no underlying climate target; instead, they model a development that follows current trends. The climate targets are not achieved in these scenarios, and they stand out from the other scenarios. In contrast, the study by Prognos and \"Oko-Institut and Wuppertal-Institut \cite{Prognos.2020b} and the survey by Fraunhofer ISE \cite{FraunhoferISE2020a} model energy scenarios in which climate neutrality is achieved that would also correspond to the actual climate protection targets in Germany (Agora KN2050, Agora KNmin, ISE Reference100). Note, that the target year to reach climate neutrality in Germany recently was shifted from 2050 to 2045 .

\begin{table}[ht!]
\caption{Overview of the reviewed energy system studies. In total, 12 studies were selected, and 37 scenarios were systematically evaluated. The selection of the studies was based on previously defined selection criteria.}

\centering
\scriptsize

\begin{tabularx}{\linewidth}{p{0.3\linewidth}p{0.25\linewidth}p{0.1\linewidth}p{0.25\linewidth}}
\toprule
System study & Scenario name & Scenario type & Emissions reduction target \\
\midrule

Dena \& ewi (2018) \cite{dena.2018a}       & 
(1) dena Baseline \newline (2) dena EL80 \newline (3) dena EL95 \newline (4) dena TM80 \newline (5) dena TM95 &  
BAU \newline Target \newline Target \newline Target \newline Target  &   
n/a \newline 80\% until 2050  \newline 95\% until 2050  \newline 80\% until 2050 \newline 95\% until 2050   \\

Prognos and \"Oko-Institut and Wuppertal-Institut (2020) \cite{Prognos.2020b}    & 
(6) Agora KN2050 \newline (7) Agora KNmin  &  
Target \newline Target & 
100\% until 2050 \newline 100\% until 2050  \\

BCG and Prognos (2018) \cite{BCG.2018}  & 
(8) BDI Reference \newline (9) BDI 80\% climate path \newline (10) BDI 95\% climate path  & 
BAU \newline Target \newline Target  & 
n/a  \newline 80\% until 2050  \newline 95\% until 2050 \\ 

LBST (2019) \cite{LBST.2019}  & 
(11) LBST EL-55\% \newline (12) LBST H2-55\% \newline (13) LBST EL-80\% \newline (14) LBST H2-80\% \newline (15) LBST EL-95\% \newline (16) LBST H2-95\%  & 
Target \newline Target \newline Target \newline Target \newline Target \newline Target  & 
55\% until 2030 \newline 55\% until 2030 \newline 80\% until 2050 \newline 80\% until 2050 \newline 95\% until 2050 \newline 95\% until 2050\\ 

ewi and ef.Ruhr (2018) \cite{ewi.2018}  & 
(17) ewi RF \newline (18) ewi EEV \newline (19) ewi TO & 
BAU \newline Target \newline Target & 
n/a \newline 80\% until 2050 \newline 80\% until 2050  \\ 

NOW (2018) \cite{NOW.2018}  & 
(20) NOW S85 \newline (21) NOW S890 \newline (22) NOW S95 & 
Target \newline Target \newline Target & 
85\% until 2050 \newline 90\% until 2050 \newline 95\% until 2050 \\ 
 
FZ Jülich (2020) \cite{Juelich.2020}  & 
(23) FZ Jül Scenario 80 \newline (24) FZ Jül Scenario 95 & 
Target \newline Target & 
80\% until 2050 \newline 95\% until 2050  \\ 

GWS and Fraunhofer ISI and DIW and DLR and Prognos (2018) \cite{GWS.2018}  & 
(25) GWS EWS & 
Target  & 
80--85\% until 2050  \\ 

Fraunhofer ISE (2020) \cite{FraunhoferISE2020a}  & 
(26) ISE Reference \newline (27) ISE Inertia \newline (28) ISE Unacceptance \newline (29) ISE Sufficiency \newline (30) ISE Reference100 & 
Target \newline Target \newline Target \newline Target \newline Target & 
95\% until 2050 \newline 95\% until 2050  \newline 95\% until 2050  \newline 95\% until 2050  \newline 100\% until 2050 \\ 

Frontier Economics et al. (2017) \cite{frontiereconomics.2017}  & 
(31) frontier EL only \newline (32) frontier EL \& gas storage \newline (33) frontier EL \& green gas & 
Target \newline Target \newline Target & 
95\% until 2050 \newline 95\% until 2050  \newline 95\% until 2050    \\ 

Prognos and Fraunhofer UMSICHT and DBFZ (2018) \cite{Prognos.2018}  & 
(34) Prognos PtX 80 \newline (35) Prognos PtX 95 &  
Target \newline Target & 
80\% until 2050 \newline 95\% until 2050  \\ 

Fraunhofer ISI and ISE (2019) \cite{Fraunhofer2019}  & 
(36) FRH Scenario A \newline (37) FRH Scenario B & 
n/a \newline n/a  & 
n/a \newline n/a  \\

\bottomrule
\end{tabularx}
\label{tab:studies} 
\end{table}

As far as the studies allowed, we extracted various data points from each scenario for 2030 and 2050. In part, it was not easy to obtain the respective data, as numbers were sometimes not explicitly mentioned, and only illustrations were shown. Nevertheless, the numbers were extracted as accurately as possible. Furthermore, different data specifications between the studies complicated the data retrieval, e.g., the question of whether gross or net data was provided or whether hydrogen was reported separately or counted as hydrogen-based carrier or even counted as \acrshort{ptx}. The explicit values used for the calculations are provided in our meta-database in the Supplementary Material.


\subsection{Data analysis}\label{sec:dataanalysis}
\label{sec:futureenergysystem}
Our analysis of the collected data of the scenarios were threefold (c.f. Section \ref{S:4}). First, to determine a more robust picture of the potential future energy system in the target years 2030 and 2050, we focused on indicators that form their main characteristics. In line with Szarka et al. \cite{Szarka.2017} and Wiese et al. \cite{wiese2022strategies} we collected and analysed the following important indicators: emissions, primary energy consumption, the share of renewables in the primary energy consumption, final energy consumption, electricity demand and percentage of renewable electricity generation. This helped us to get more insights into the assumptions and the differences of the various scenarios as presented in Section \ref{S:41}. Furthermore, we were able to define deviations and ranges of the various studies for the corresponding parameters.

Second, the scenarios were compared regarding the future role of hydrogen and hydrogen-based synthetic fuels in 2030 and 2050. For this, the study results are discussed concerning the following indicators: hydrogen demand, synthetic fuels demand and the composite indicator \acrshort{ptx} demand, which summarises the demand for both hydrogen and synthetic fuels. Ignoring synthetic energy carriers, which are often based on hydrogen, would distort the results of the respective indicators. Based on these findings, the domestically produced and imported rate of \acrshort{ptx} products were determined. Furthermore, the importance of \acrshort{ptx} in the energy system was analysed with the help of the variable share of \acrshort{ptx} in final energy consumption, which was put into association with the decarbonisation targets of the studies. This data collection and assessment is according to our knowledge also unique and is presented in Section \ref{S:42}. Additionally, we also calculated the import quotas and the domestic production possibilities of hydrogen and other synthetic fuels according to the reviewed scenarios.

Third, we subsequently determined the value creation according to the numbers in the studies and scenarios since hydrogen and synthetic hydrogen-based products offer the potential for domestic economic benefits. Value creation in this paper is defined as “a share of the system costs that are provided by companies in Germany” \cite{LBST.2019}. We adopted the calculation approach of Wuppertal Institute and DIW Econ \cite{WuppertalInstitut.2020}, which is based on the previously published method and values of LBST \cite{LBST.2019}. The latter calculation approach is purely cost-based and takes into account the costs of the plants as well as the wages paid by companies. Profits and margins, however, are not considered \cite{LBST.2019}. 

Analogous to the \cite{WuppertalInstitut.2020} calculations, the assumptions regarding techno-economic parameters, such as the capacity of plants, investment costs and employment intensities are adopted from the study by LBST \cite{LBST.2019}. Hence, it is also assumed that hydrogen is produced exclusively from electrolysis and that electricity originates from renewable sources. In contrast to the previous studies \cite{LBST.2019,WuppertalInstitut.2020}, this paper examined not only the value created by hydrogen but also by other \acrshort{ptx} products such as synthetic energy carriers. For ease of calculation, all \acrshort{ptx} products have been treated as hydrogen \footnote{We are aware that this view neglects that synthetic fuels can also be produced with other products than hydrogen, e.g., biomass. Moreover, the economic effects resulting from the further processing of hydrogen are not taken into account. Furthermore, non-hydrogen \acrshort{ptx} products are treated as if they could be stored and transported like hydrogen. This ignores the fact that products like \acrshort{ptl} have different properties than hydrogen and that storage and transport of these fuels will therefore be carried out differently.}. The hydrogen value chain for which the effects are examined consists of four stages: (1) renewable electricity generation that is needed for the second part; (2) the production of hydrogen via electrolysis; (3) hydrogen storage in salt caverns and pipe storage facilities is the third part, followed by (4) the hydrogen transport which includes the conversion of natural gas pipelines to hydrogen pipelines and the operation of these pipelines. The economic effects of end-use applications are not considered in this paper. For each of these four levels, the economic effects can be calculated separately setting certain assumptions as outlined in Section \ref{S:43}. 

Like in \cite{WuppertalInstitut.2020}, we also used the EL-55\% 2030 scenario\footnote{While "EL" marks studies or rather scenarios with focus on electrification, H2 marks studies or rather scenarios with a focus on the use of hydrogen} \cite{LBST.2019} as basis for calculating the value creation effects for 2030. The the EL-80\% 2050 scenario was used as the basis for the effects for 2050. In contrast, the hydrogen transport as fourth step of the value chain is based on scenario H2-80\% for 2030 and 2050. This is because in the electrification scenarios it is generally assumed that the electrolysis plants are installed as close to the consumer as possible (e.g., at petrol stations), so hydrogen transport routes are no significant factor \cite{LBST.2019}. While assuming a linear relationship of the given economic values and the given \acrshort{ptx} demand, (1) we calculated the value creation of the first and second stage (production) as product of the national production share of \acrshort{ptx} and the value creation effect for one given unit of the basis scenarios. In contrast, the value creation of stage three and four were calculated with the whole \acrshort{ptx} demand and not only with the domestic production share as these steps are required regardless of whether the hydrogen is produced domestically or abroad \cite{WuppertalInstitut.2020}. The calculations are outlined in the Supplementary Material.

\section{Future role and value creation of hydrogen and synthetic fuels}
\label{S:4}

\subsection{Assumptions and pathways of future long term energy system scenarios}
\label{S:41}

Many studies use integrated models of the energy system to assess future developments. Assumptions are made that may differ among the studies. An overview of the determined variables and parameters of the 37 scenarios is provided in Figure \ref{fig:StudiesOverview}. As described in the methodology section, only studies that provide the required data are included in the following analyses. Moreover, some studies focus on the year 2050 and do not supply information for 2030.

\begin{figure}[ht!]
    \centering
    \includegraphics[width=0.7\textwidth]{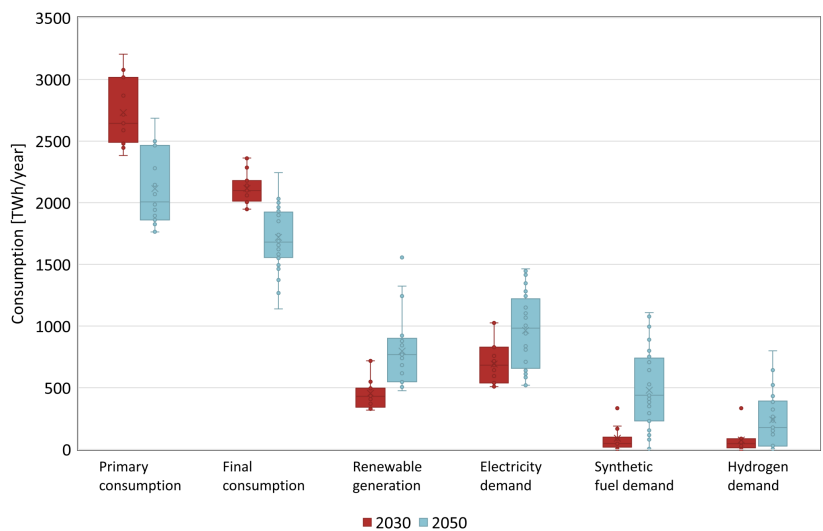}
    \caption{Overview of main characteristics (assumptions and calculations) (c.f., Section \ref{sec:futureenergysystem}) of energy system scenarios when modelling future German transition pathways for the year 2030 and 2050. The boxplots are indicating the values from the 25\%-75\% quartile. While the median is indicated with the solid line, the mean is highlighted with the marker.  Own depiction based on the collected study data. Each point can be associated to one of the 37 scenarios of the twelve studies as presented in Table \ref{tab:studies}.}
    \label{fig:StudiesOverview}
\end{figure}

The current primary energy consumption of 3,248 TWh (or 11,691 PJ \cite{UBA.2021b}) is expected to decline by the scenarios in 2030 (2,383 TWh (Agora KN2050) to 3,205 TWh (BDI Reference)). In the scenarios with high climate targets (e.g., Agora KN2050 and Agora KNmin), the primary energy demand decreases at a much higher rate. By 2050, today's primary energy demand is expected to be declined by about one third to 2,114 TWh on average. At the same time, the share of renewable energies in primary energy demand is expected to increase. While \acrshort{re} production is already expected to increase by 2030, with an average value of just under 30\%, there is yet a further significant increase thereafter. By mid-century, according to the scenarios, about two-thirds of primary energy demand will come from renewable sources (mean 2030: 28\%; mean 2050: 67\%). The scenarios also estimate that the final energy consumption will decline from the current level of around 2,500 TWh. On average, the value is 2,117 TWh. By 2050, the final energy demand in the scenarios is expected to decrease to 2,245 TWh in the ISE Inertia scenario and 1,139 TWh in LBST EL-95\%. The mean is 1,712 TWh, and thus about one-third below today's value. Thereby, the share of fossil fuels decrease and the electricity share increases. 

Electricity demand is expected to increase significantly in most scenarios, especially in the long term (2020: 553 TWh; mean 2030: 692 TWh; mean 2050: 958 TWh). The lowest number is 510 TWh/a in the scenarios Prognos PtX 80 and 95 in 2030. In contrast, the highest value for 2030 is reached in scenarios LBST H2-80\% and LBST H2-95\%. For 2050, the demand for electricity in Prognos PtX 80 and 95 remains similarly low as in 2030 (521 TWh). In LBST H2-95\%, ISE Reference and ISE Inertia, however, this demand is higher than 1,400 TWh/a. It is because electricity is needed either directly for end-use applications or indirectly to produce hydrogen via electrolysis. However, there are exceptions. The decisive factor for the level of electricity demand is how much synthetic fuels are imported. In studies that foresee a high share of such imports, the electricity demand is lower. In all examined scenarios, the share of coal-fired electricity is declining significantly by 2030 or is already at zero. Instead, gas and renewable energies are becoming more critical. To achieve the climate targets, electricity will increasingly be produced from renewable energies. In 2050, the share of renewables in the electricity sector exceeds 70\% in all scenarios and thus represents by far the most important type of power generation. The mean share is 90\%, whereby dena Baseline indicates the lowest share at 72\%, while in other scenarios even a share of 100\% is reached.

\begin{figure}[ht!]
    \centering
    \subfloat[2030]{\includegraphics[width=0.7\textwidth]{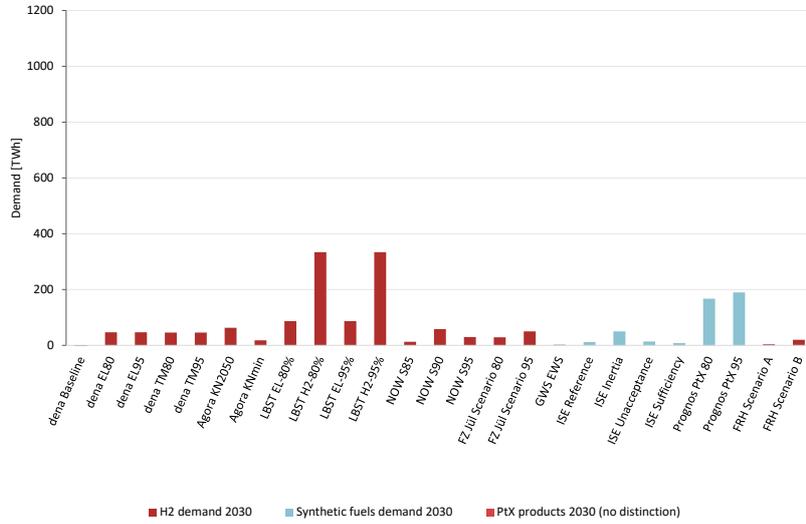}}\\
    \subfloat[2050]{\includegraphics[width=0.7\textwidth]{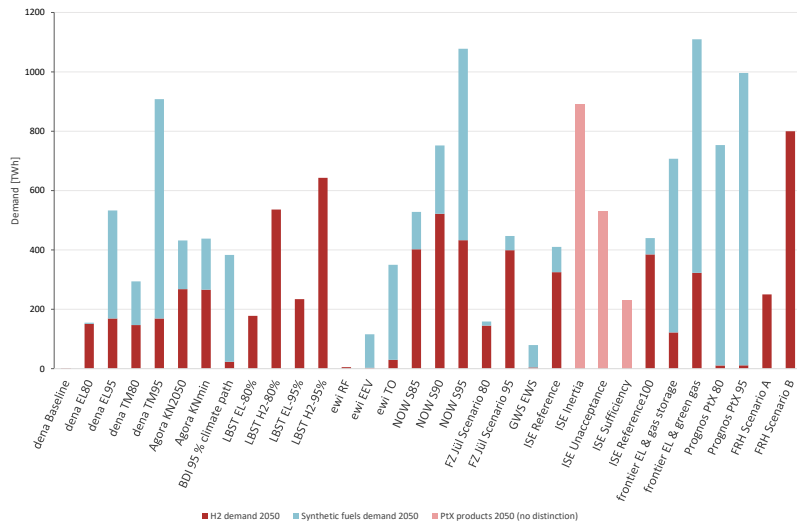}}
    \caption{Reported \acrshort{ptx} (hydrogen and synthetic fuel demand and only \acrshort{ptx} products) in 2030 and 2050. In 2030, the demand for \acrshort{ptx} products is around 80 TWh on average. Synthetic fuels other than hydrogen only play a major role in two scenarios. The demand for synthetic products will increase significantly by 2050 according to the scenarios and is on average about 480 TWh. The 37 scenarios of the twelve studies as presented in Table \ref{tab:studies}}
    \label{fig:DemandHydrogen}
\end{figure}

In terms of hydrogen, various scenarios do not foresee a broad use by 2030. The mean quantity demanded is about 60 TWh, only LBST H2-80\% and H2-95\% already see a demand of 334 TWh (c.f., \ref{fig:DemandHydrogen}). Until 2050, several scenarios expect hydrogen to play an important role in the energy system. However, they also vary in this respect. While some scenarios expect no hydrogen use at all, many see the demand between 100 and 200 TWh, FRH Scenario B even at 800 TWh. The reported mean demand in 2050 is 240 TWh. At the same time, scenarios that expect comparatively low hydrogen demand such as Prognos PtX 80 and 95, BDI 95\% climate path, or ewi EEV and ewi TO, project a significant demand for other synthetic energy carriers. On average, about 320 TWh of synthetic fuels are demanded in the scenarios. This number is higher than the demand for hydrogen, although it should be noted that hydrogen is often needed to produce \acrshort{ptx} fuels. For 2030, no notable demand for other synthetic fuels (excluding hydrogen) is assumed in any scenario except frontier EL \& gas storage and frontier EL \& green gas (167 TWh and 189 TWh, respectively).  

Thus, most studies also do not expect a widespread introduction of \acrshort{ptx} products (hydrogen and synthetic carriers) in 2030. The exceptions are the LBST H2-80\% and H2-95\% scenarios, which assume demand for hydrogen of 334 TWh, and Prognos PtX 80 and 95, which predict demand for synthetic products of 167 TWh and 190 TWh, respectively. These last two are also the only ones among all scenarios that see any demand for synthetic fuels in 2030. A reason for this could lie in the assumptions and the client of the studies. Both studies requests for an explicit use of the carriers in different sectors. Furthermore, the client for the Prognos study was the mineral oil industry. On average, 79 TWh of synthetic energy sources (including hydrogen) are demanded. In contrast, in almost every scenario in 2050, either hydrogen or other synthetic energy carriers or both, respectively, will be used. \acrshort{ptx} products are not used in the dena Baseline scenario and are hardly used in the ewi RF scenario (5 TWh). In these scenarios, however, the climate targets are not achieved either. GWS EWS also has restrictive assumptions regarding the use of \acrshort{ptx} and assumes a demand of 79 TWh, of which the majority are synthetic products. The highest quantities of \acrshort{ptx} products are projected in frontier EL \& green gas (1,110 TWh) and NOW S95 (1,078 TWh). In Prognos PtX 95, demand almost reaches 1,000 TWh. The combined hydrogen and hydrogen-based synthetic fuel demands for each study is summarised for 2030 and 2050 in Figure \ref{fig:DemandHydrogen}.

\newpage
\subsection{Importance of hydrogen and synthetic fuels with respect to long term scenarios}
\label{S:42}
\begin{figure}[ht!]
    \centering
    \includegraphics[width=0.7\textwidth]{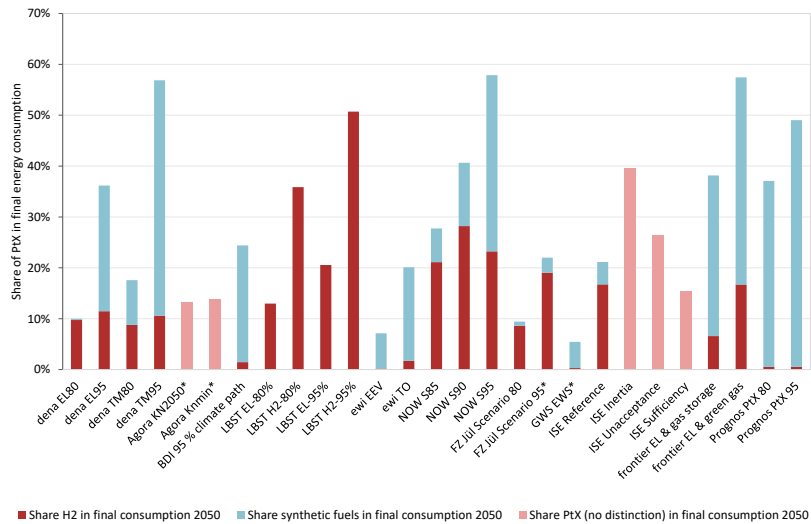}
    \caption{Share of \acrshort{ptx} in final energy consumption for the year 2050. By 2050, hydrogen and other synthetic fuels will have an average share of almost one quarter in final energy consumption. However, the scenarios differ significantly from each other. The share of non-hydrogen synthetic energy carriers is larger than the share of hydrogen. In the scenarios marked with `*', the share of \acrshort{ptx} in final energy demand was reported separately.} 
    \label{fig:sharefinalptx}
\end{figure}

The importance of hydrogen and other \acrshort{ptx} products for the overall system can be initially assessed with the share of hydrogen in the final energy demand. An overview of the shares in each study for 2050 is provided in Figure \ref{fig:sharefinalptx}.\footnote{Due to a lack of more precise data, the \acrshort{ptx} share in the final energy demand was calculated from the demand for hydrogen and synthetic fuels, i.e., including the demand of the electricity sector. When \acrshort{ptx} is used for reconversion, the amount of \acrshort{ptx} converted into electricity no longer counts as final energy. Therefore, the indicator is inaccurate and only represents an approximation of the real value. Only in the scenarios marked with `*', the share of \acrshort{ptx} in final energy consumption was reported separately for the scenarios.} 

By 2030, the two LBST H2 scenarios and the two Prognos scenarios stand out. While in LBST H2-80\% and H2-95\% the share of hydrogen in the final energy demand in 2030 is 16\%, in the Prognos scenarios, the share of synthetic fuels is 4\% and 6\%, respectively. The other scenarios do not expect a significant percentage of \acrshort{ptx}. The figure thus confirms the impression that hydrogen and other \acrshort{ptx} products will probably not play a significant role until 2030, the mean share in final energy consumption is 4\%. However, in many studies, hydrogen will have already started to be integrated into the energy system. Synthetic fuels, on the other hand, are only expected in significant quantities by Prognos, Fraunhofer UMSICHT and DBFZ \cite{Prognos.2018}.

\begin{figure}[ht!]
    \centering
    \includegraphics[width=0.7\textwidth]{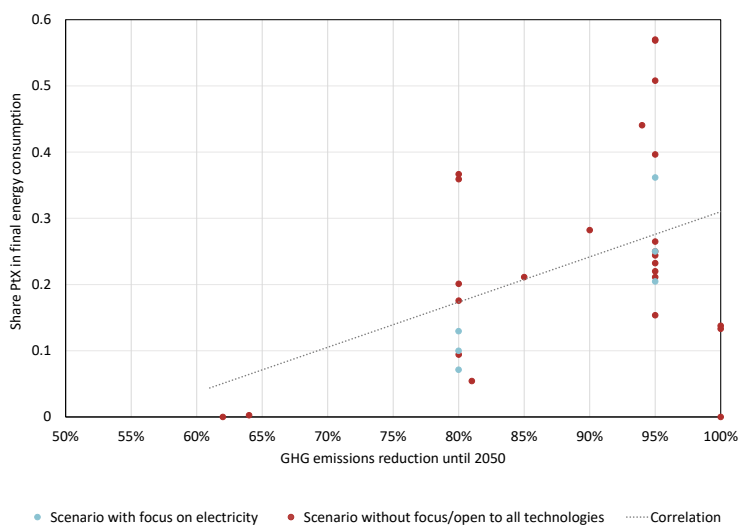}
    \caption{Correlation between the decarbonisation target and the share of \acrshort{ptx} in final energy consumption in 2050. The figure indicates a positive correlation between the reduction of GHG emissions and the share of \acrshort{ptx} in final energy consumption. The correlation coefficient is 0.53. Each point can be associated to one of the 37 scenarios of the twelve studies as presented in Table \ref{tab:studies}.} 
    \label{fig:correlation}
\end{figure}

In 2050, the picture is different. Among the scenarios, only GWS EWS (5\%) and ewi EEV (7\%) indicate a share of \acrshort{ptx} products in final energy demand of less than 10\%. Four studies even report more than 50\% share, the highest number being 58\% in NOW S95. By 2050, the share of \acrshort{ptx} in final energy demand is on average 24\% (c.f. Figure \ref{fig:sharefinalptx}). Even if this value is more of an upper limit due to the neglect of reconversion, it demonstrates that hydrogen and synthetic fuels will be an important component of the future energy system. Depending on the respective assumptions, the focus in some scenarios is more on hydrogen (as in the LBST scenarios) and in others on synthetic fuels (as in the Prognos scenarios). In most scenarios, both hydrogen and hydrogen-based energy carriers such as \acrshort{ptx} and \acrshort{ptg} are used. Non-hydrogen synthetic energy carriers account for a larger share than hydrogen. Significant \acrshort{ptx} shares are even achieved in dena EL80, dena EL95, LBST EL-80\% and EL-95\% or ewi EEV when aiming for the electrification of the system. 
Besides the choice of the technological focus of the scenarios, the climate target modelling has the most significant influence on the level of the \acrshort{ptx} share. This can be seen particularly well in scenarios of the same study with different GHG emission reduction targets, for which the same assumptions apply in each case. A tightening of the climate target from 80\% to 95\% GHG emissions reduction by 2050 compared to 1990 leads to a tripling of the \acrshort{ptx} share in the scenarios in \cite{dena.2018a}. The increase between scenarios is not as substantial, but still noticeable within other studies. 
This observation leads to the conclusion that the decarbonisation of the energy system and the share of PtX in it are interrelated, which is illustrated by Figure \ref{fig:correlation}. 
The figure confirms the previous assumption that hydrogen and other hydrogen-based products (summarised with PtX) are positively associated with the decarbonisation targets. 

Finally, there are also apparent differences in domestic hydrogen and \acrshort{ptx} production share. In 2030, when hydrogen demand is estimated to be relatively low, the scenarios assume at least 30\% (Agora KN2050) domestic production. Many scenarios consider that the entire demand can be produced domestically, the average share being over 80\%. With higher hydrogen demand in 2050, the picture almost splits in two. While one group of studies considers domestic production shares of 80\% to 100\% to be possible, the other group expects shares of only roughly one third. Two studies are between, estimating about 65\%, which is about the average value (67\%). The ewi RF scenario expects the projected very low hydrogen demand of 5 TWh to be imported. Some studies do not provide precise data on where the hydrogen is produced. In the LBST and frontier scenarios, e.g., it is also assumed that hydrogen is only produced domestically \cite{LBST.2019}.
In contrast, other scenarios also considers hydrogen production by steam reforming or bio-to-H2 as well as imports. Additionally, synthetic fuels are assumed to be mostly imported. ISE Reference is the only scenario that expects more than half of the synthetic fuels demand to be produced domestically. On average, domestic production accounts for about 10\% of the quantity demanded. Prognos and NOW do not provide explicit data on imports, but both studies emphasise the opportunities of synthetic fuel imports.

\subsection{Value creation effects of hydrogen and synthetic fuels}
\label{S:43}
\begin{figure}[ht!]
    \centering
    \includegraphics[width=0.7\textwidth]{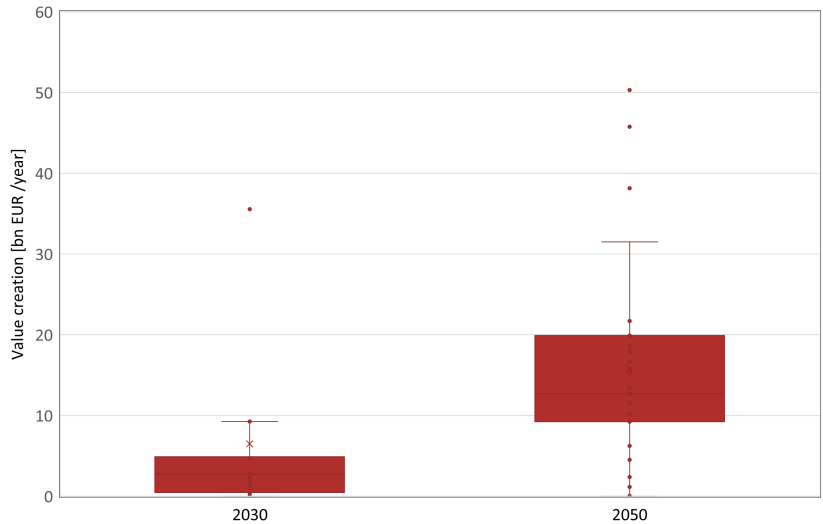}
    \caption{Value creation by the use of hydrogen and hydrogen-based carriers in the year 2030 and 2050. While the value creation effects are about 4.5 bn EUR in 2030, they are on average 15.8 bn EUR for the year 2050. Each point can be associated to the results of one of the 37 scenarios of the twelve studies as presented in Table \ref{tab:studies}.}
    \label{fig:valuecreation}
\end{figure}

\begin{figure}[ht!]
    \centering
    \subfloat[2030]{\includegraphics[width=0.7\textwidth]{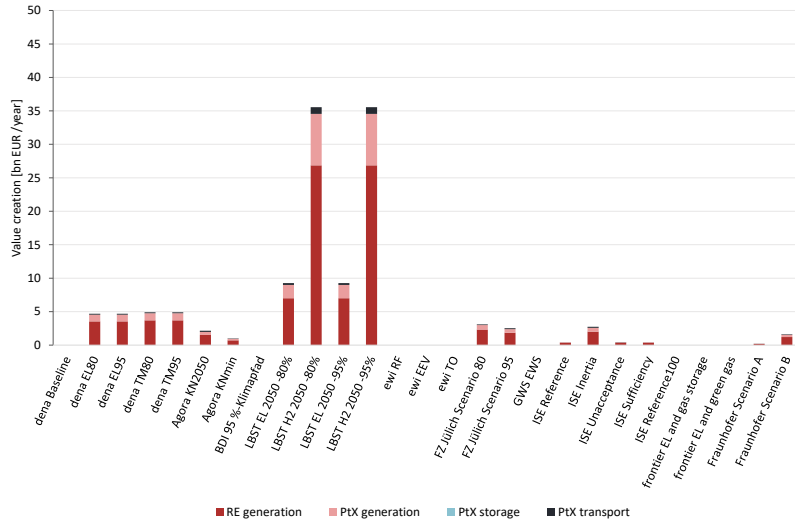}}\\
    \subfloat[2050]{\includegraphics[width=0.7\textwidth]{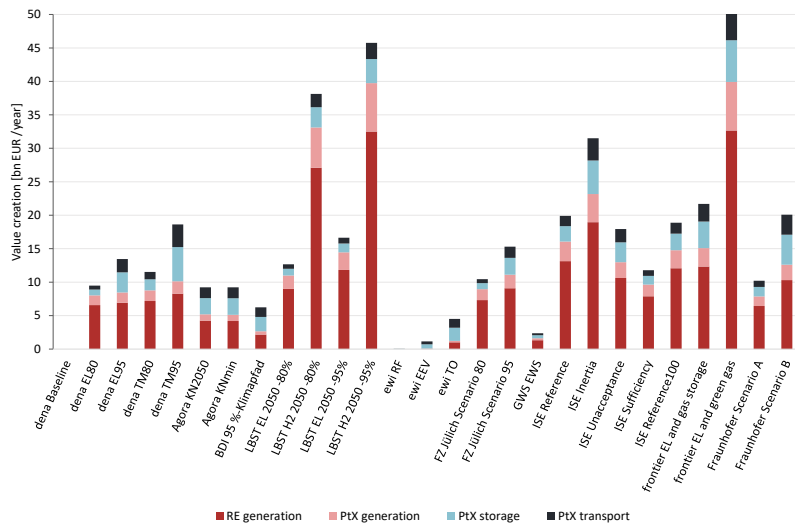}}
    \caption{Value creation in 2030 and 2050 by the use of \acrshort{ptx}  with respect to the individual value chains. The value creation effects of hydrogen and \acrshort{ptx} are about 4.5 bn EUR. Synthetic fuels are hardly used in the scenarios by 2030. The value creation effects by the implementation of hydrogen increases to 12.3 bn EUR on average in the year 2050. Including other synthetic fuels, the average value added grows to 15.8 bn EUR. The largest share of value creation is generated in the sector of renewable electricity generation. The 37 scenarios of the twelve studies as presented in Table \ref{tab:studies}}
    \label{fig:valuechain}
\end{figure}

The potential value creation effects of hydrogen and hydrogen-based carriers in Germany in the years 2030 and 2050 are outlined in Figure \ref{fig:valuecreation}. The value for each study is presented with respect to the different steps of the value chain in Figure \ref{fig:valuechain}. As described above, the calculation is based on the projected fuel demand and the expected share of domestic production in the respective scenarios. Furthermore, only studies that provide the required data for the years 2030 or 2050 are included in the calculation. 

According to the majority of the scenarios, the value created in 2030 ranges from 0 to 5 bn EUR/a. Only the LBST scenarios stick out with 9.3 bn EUR/a (EL scenarios) and 35.6 bn EUR/a (H2 scenarios). By 2050, most scenarios expect hydrogen and/or synthetic fuels to be an important part of the energy system. The use and production of hydrogen or hydrogen-based products generate significant value-added effects, although the results differ considerably between the scenarios. The value creation effects are almost 10 bn EUR/a higher in scenario LBST H2 2050 -95\% than in 2030, which is four times higher than the increase in scenario LBST H2 2050 -80\%. Only four of the depicted scenarios do not anticipate any or nearly any value creation effects. Similar to 2030, the highest value added effects are given by the two LBST H2-scenarios with 38 and almost 46 bn EUR/a, respectively. 

Overall, the scenarios predict significant value creation in 2050. The difference compared to the results for 2030 where almost no non-hydrogen synthetic fuels are expected to be used is particularly large. In 2050, nearly all scenarios expect synthetic energy carriers to be implemented in the energy system which leads to value creation. No value creation effects are identified only in the BAU scenarios. The reason is that these studies are using majorly energy sources which are currently available. The scenarios ewi EEV and GWS EWS show very low values with 1.1 bn EUR/a and 2.4 bn EUR/a, respectively, while the frontier EL and green gas scenario (50.3 bn EUR/a) even reaches a higher value added than Scenario LBST H2 2050-95\%. 

Regarding the different stages of the value chain, the Figure \ref{fig:valuechain} shows that renewable electricity generation has by far the largest share in value creation. Even if the proportion fluctuates between the scenarios, it accounts between 50\% and 70\% for the majority of scenarios. In the calculation, it is assumed that the electricity is produced entirely in Germany.
The hydrogen or synthetic fuels production accounts for the second-highest share in most scenarios, whereby it depends on the proportion that is produced domestically. The value creation effects from transport and storage are independent of this share, they are only related to the respective demand of the studies. We assume that we need the transport and storage facilities regardless of whether it is produced domestically or internationally. If, as in the BDI 95\% climate path or ewi TO scenario, there is a strong demand for hydrogen or \acrshort{ptx} but the domestically produced share is low, higher shares of the transport and storage sector in gross value added occur.

\section{Discussion}
\label{S:5}

\subsection{Key findings and implications}
The German energy system is expected to change significantly according to the analysed scenarios. All scenarios expect a significantly reduced energy demand by 2050, which is reflected in both lower primary energy consumption and final energy consumption. The share of renewable energies in primary energy consumption increases from less than 20\% today to at least 50\% in target scenarios. Although hydrogen and synthetic fuels will play an important role according to the collected scenarios, by 2030 the demand of carbon-neutral hydrogen is still moderate with approximately 50 TWh. The LBST-H2 scenarios, which foresee a demand of over 300 TWh already in 2030, can be considered an exception. While the demand for hydrogen is low but present in most scenarios until 2030, the demand for synthetic fuels only exists in one study or two scenarios. 

The identified hydrogen demand for 2030 coincides with the literature that partly examine the same scenarios as those considered in this paper \cite{WuppertalInstitut.2020,adelphi.2019} . Concerning the German government's hydrogen demand target of 90 to 110 TWh in 2030 which is specified in the national hydrogen strategy \cite{BMWi.2020}, it can be detained that most scenarios remain below it. Therefore, this goal can be considered very ambitious and additional efforts might be needed to realise it. However, the expected domestic production of up to 14 TWh of hydrogen is exceeded in nearly all scenarios. The LBST-H2 scenarios \cite{LBST.2019}, which foresee a demand of over 300 TWh already in 2030, can be considered an exception. It is noticeable that many scenarios assume that the demand for hydrogen in 2030 can be met by domestic production. While the demand for hydrogen is low but yet present in most scenarios until 2030, the demand for synthetic fuels only exists in one study or two scenarios, respectively. In these scenarios, however, it is already comparatively high at over 150 TWh. With regard to the German government’s hydrogen demand target of 90 to 110 TWh in 2030 which is specified in the national hydrogen strategy \cite{BMWi.2020}, it can be stated that most scenarios remain below it. This goal can therefore be considered very ambitious and additional efforts might be needed to realise it. At the same time, the strategical domestic production of  14 TWh is exceeded in all scenarios that already see such a demand for hydrogen in 2030.

By 2050, most scenarios project a significant increase in demand for hydrogen. However, the figures vary widely, with a demand of up to 800 TWh. In some scenarios with comparatively low demand for hydrogen instead, a significant demand for synthetic fuels other than hydrogen is assumed. The total demand for \acrshort{ptx} products as well as the allocation between hydrogen and synthetic fuels within the scenarios differ substantially between the projections. The mean demand in 2050 of the examined scenarios is 480 TWh, which is about six times higher than in 2030. This development indicates that electricity conversion into \acrshort{ptx} products is expected to be a substantial part of the future Germany energy system. Another aspect that is emphasised by the study of Adelphi, Wuppertal Institute, and IEEJ \cite{adelphi.2019} is the positive correlation between climate protection targets and the relevance of \acrshort{ptx} in the energy system. This statement can be confirmed. The correlation between the level of \acrshort{ghg} reduction by 2050 and the share of \acrshort{ptx} in final energy consumption is positive. In scenarios that assume the achievement of stricter climate targets, such as an emissions reduction of 95\% instead of 80\% by 2050, the \acrshort{ptx} demand is higher. The same applies to scenarios that assume openness to technology instead of a focus on electrification. These conclusions seem logical: Increased climate protection targets result in more conventional energy sources being replaced by synthetic ones. If the focus is on electrification, in turn, electricity will be used more often in end-use applications than synthetic products, resulting in lower \acrshort{ptx} demand.Nevertheless, the share of corresponding energy sources in final energy consumption varies in the target scenarios between 5\% and 57\%, with an average of almost one-quarter (24\%). While many scenarios assume that hydrogen can be produced domestically to a significant extent (67\%), other synthetic products are often expected to be imported to a major proportion (on average only 8\% are produced in Germany). The high variance in synthetic fuel import was also mentioned by Wiese et al. \cite{wiese2022strategies}.

Since synthetic products are often based on hydrogen, the value creation effects of the use of hydrogen and its derivatives were also taken into account in this paper. For the calculation, the demand for \acrshort{ptx} products was considered while the effects of the conversion of hydrogen into synthetic products were neglected. Even though the use of hydrogen and other synthetic fuels will have only just begun at 2030 and will increase significantly by 2050, the determined economic value is already more than half of the amount of the federal government's funding. By 2050, the value-added effects from the use of hydrogen in the energy system is expected to almost triple compared to 2030, with a mean value of more than 12 bn EUR/a. When also adding the effects from using other synthetic energy carriers, the average is almost 16 bn EUR/a. This is approximately 15\% of the gross value added of the German automotive industry in 2018 \cite{Statista}. A comparison with the results of LBST \cite{LBST.2019} show that the calculations for 2030 are almost identical. Also, for 2050, the LBST -80\% scenarios still show similar results, while the results of the LBST -95\% scenarios differ. This deviation can be traced back to the fact that in the respective scenario in LBST \cite{LBST.2019}, an additional electrolysis capacity is assumed to be installed. This capacity not only serves to cover the hydrogen demand but is also used as a flexible load to stabilise the energy system due to the high RE shares. This effect is not covered in the methodology of this paper since the role of electrolysis capacity for load stabilisation is not taken into account.

In line with other existing publications \cite{LBST.2019,WuppertalInstitut.2020} a large part of the value creation can be attributed to the generation of renewable electricity. Thus, renewable electricity generation is not only a prerequisite for a successful energy transition but also offers high potential for economic benefits. Besides, the highest values are achieved in scenarios that assume that the entire production of \acrshort{ptx} products occurs domestically. While the government's hydrogen strategy considers the North Sea region and southern Europe as suitable regions from which hydrogen could be imported \cite{BMWi.2020}, some researchers argue that the imports of green hydrogen from regions abroad are not necessarily cheaper than a production in Germany or at least a too high import rate is not optimal \cite{WuppertalInstitut.2020}. Despite of the necessary and high investments to build the facilities, a domestic production could have favourable impacts. Additionally, the share of \acrshort{ptx} in final energy consumption of Germany in 2050 (24\%) seems comparable but a little bit to the mean European share and also way higher than the global share. Wiese et al. \cite{wiese2022strategies} even questions whether the assumed imported amounts of hydrogen and synthetic fuels can actually be produced outside Germany in a carbon-neutral way.

\subsection{Limitations of the chosen approach}

The presented results need to be critically viewed. On the one hand, the collected data basis is quite heterogeneous. While the studies applied different foci, they are also founded on different energy system assumptions and data. Furthermore, the data collection obtained from the studies was not always straightforward. The studies used different ways of presenting their results, some of them only showing relevant data in graphs without explicitly stating the values. The values were estimated as accurately as possible in these cases, but small deviations may have occurred.

Further limitations concern the calculation of economic effects which might be overestimated. Since we used the same calculation method as stated by Wuppertal Institute and DIW Econ \cite{WuppertalInstitut.2020}, we also rely on similar drawbacks. The computation was based exclusively on the demand for hydrogen or \acrshort{ptx}, respectively, as well as the share of domestic production while assuming that the value-creating resulting from domestic hydrogen production is linear \cite{LBST.2019}. Thereby, e.g., the attribution of the economic effects from the expansion of \acrshort{re} to the hydrogen-related products is discussionable since renewable electricity is also required in alternative scenarios. The assumption that hydrogen is exclusively generated by water electrolysis is also quite restrictive since, e.g., the scenario 80 in the study of FZ J\"ulich \cite{Juelich.2020} states that about two-thirds of the required hydrogen is produced by steam methane reforming. 

However, some aspects also suggest that the results might be underestimated. Economic effects from the use of \acrshort{ptx}, e.g., were neglected in the study of Wuppertal Institute and DIW Econ \cite{WuppertalInstitut.2020}, as was the export of plants and equipment that could result from the global demand for hydrogen. Against the background of the strong export orientation of German machinery and plant manufacturers, there is significant economic potential in this sector \cite{WuppertalInstitut.2020,frontiereconomics.2018}. In contrast to LBST \cite{LBST.2019}, and Wuppertal Institute and DIW Econ \cite{WuppertalInstitut.2020}, the calculation of the economic effects in this paper also included other \acrshort{ptx} products. At the same time, the economic effects from the conversion of hydrogen into synthetic fuels like \acrshort{ptl} were neglected.

Additionally, one should keep in mind that the economic effects in LBST \cite{LBST.2019} are based on a “self-contained and integrated energy system in which the optimisation of capacities and all system elements, including hydrogen, occurs simultaneously” \cite{WuppertalInstitut.2020}. Changes in the components of some elements of this system, such as the demand for hydrogen or the share of domestic production, would possibly lead to the adjustment of constraints and a change in the economic potentials. Nevertheless, according to the Wuppertal Institute and DIW Econ \cite{WuppertalInstitut.2020}, the determined effects would provide an overall approximation of the economic potential of a domestic hydrogen production differently developed in the future.

\section{Conclusion}
\label{S:6}

Hydrogen and synthetic fuels will play an important role in the future energy system. Among the scenarios examined, there is none in which the climate targets are achieved without the use of hydrogen or other synthetic energy carriers. According to the scenarios, the use of hydrogen can be expected to have started by 2030. The demand for hydrogen in 2030 is around 50 TWh per year (\textasciitilde4\% of the final energy demand) in many scenarios, while with few exceptions other synthetic fuels are not expected to be used. Consequently, in most scenarios, the federal government's hydrogen demand target of 90 to 110 TWh that is specified in its hydrogen strategy will not be reached. At the same time, the expected domestic production of up to 14 TWh of hydrogen in 2030 is exceeded in various scenarios. By 2050, hydrogen and other synthetic energy products will be an essential part of the energy system according to the scenarios, with a mean demand of 480 TWh. On average, these energy carriers account for about a quarter of final energy consumption (\textasciitilde24\%). The analysis of the scenarios also shows that \acrshort{ptx} products have a higher share in final energy consumption on average in scenarios with stricter climate targets. A positive correlation (0.53) exists between the level of GHG reduction by 2050 and the share of \acrshort{ptx} in final energy consumption.

According to the calculations, an average of almost 5 bn EUR per year in value is added by \acrshort{ptx} in German. In 2050, the value goes up tp 16 bn EUR per year. The implementation of the energy transition, which is necessary to comply with the Paris Climate Agreement, and the development of domestic hydrogen production depends to a large extent on the expansion of \acrshort{re}. Policymakers should therefore pay particular attention to the rapid expansion of \acrshort{re}. Despite of the various advantages, one always has to keep in mind that the construction of the facilities for the carriers is associated with high costs. Thus, it would be necessary to deduct investment costs, planning and installation costs, and operational costs for each value creation stage to get the final benefit. Thus, we believe that the conclusions and pathways encountered in this meta-analysis about the future German system are also transferable to other demand-intensive countries in Europe and worldwide.

Overall, it can be concluded that hydrogen and other \acrshort{ptx} products will not only contribute to achieving climate targets but will also provide economic benefits. Given the fact that most studies predict hydrogen derivatives to be largely imported in the future, investigating the economic effects of converting hydrogen into synthetic energy carriers could be an interesting approach for further research. Since the avoided environmental damage alone relieves society of enormous costs, a more in-depth analysis regarding the exact proportion of hydrogen and its further economic benefits should be conducted in further research. Additionally, with an increasing demand worldwide for hydrogen and synthetic fuels a better assessment and comparison of the national import and export quotas is necessary.

\section*{Supplementary material} 
The Supplementary Material (SM) consists of the scenario database. All extracted data points of the 12 studies and 37 scenarios are collected there. Furthermore, the calculations for the value creation are presented.

\section*{Declaration of competing interest} 
The authors declare that they have no known competing financial interests or personal relationships that could have appeared to influence the work reported in this paper.

\section*{Acknowledgement}
This research is funded by the German Federal Ministry of Education and Research (BMBF) by the project “Zwanzig20 - HYPOS - Joint project storage study” with the project number 03ZZ0753.




\newpage
\bibliographystyle{elsarticle-num}
\bibliography{ModelDescription.bib}







\end{document}